\documentstyle[11pt,epsfig]{article}
\newcommand{\NP}[1]{{\it Nucl.\ Phys.}\ {\bf #1}}

\newcommand{\PL}[1]{{\it Phys.\ Lett.}\ {\bf #1}}
\newcommand{\PR}[1]{{\it Phys.\ Rev.}\ {\bf #1}}
\newcommand{\PRL}[1]{{\it Phys.\ Rev.\ Lett.}\ {\bf #1}}

\newcommand{\EPJ}[1]{{\it Eur.\ Phys.\ J.}\ {\bf #1}}
\newcommand{\be}{\begin{equation}}
\newcommand{\ee}{\end{equation}}
\newcommand{\bea}{\begin{eqnarray}}
\newcommand{\eea}{\end{eqnarray}}

\newcommand{\bfk}{\mbox{\boldmath $k$}}
\newcommand{\pup}{p^\uparrow}

\newcommand{\bfp}{\mbox{\boldmath $p$}}
\newcommand{\bfP}{\mbox{\boldmath $P$}} 
\newcommand{\Lup}{\Lambda^\uparrow} 
 
\newcommand{\hup}{h^\uparrow} 
\newcommand{\hdown}{h^\downarrow} 
\newcommand{\nd}{\noindent}
\renewcommand{\thefootnote}{\fnsymbol{footnote}}

\begin{document}
\pagestyle{plain} 
\setcounter{page}{1} 

\vspace*{.3cm}

\begin{center}
{\bf Transverse $\mbox{\boldmath$\Lambda$}$ polarization
 in inclusive processes\footnote{
Talk delivered by U.~D'Alesio at the 3rd Circum-Pan-Pacific Symposium on
``High Energy Spin Physics'', Beijing, China, October 8-13, 2001.}}\\
\vskip 0.6cm
{\sf M.~Anselmino$^1$, D.~Boer$^2$, U.~D'Alesio$^3$, F.~Murgia$^3$}
\vskip 0.5cm
{\it $^1$ Dipartimento di Fisica Teorica, Universit\`a di Torino and \\
          INFN, Sezione di Torino, Via P. Giuria 1, I-10125 Torino, Italy}\\
\vspace*{0.3cm}
{\it $^2$   Dept.\ of Physics and Astronomy, Vrije Universiteit Amsterdam, \\
De Boelelaan 1081, 1081 HV Amsterdam, The Netherlands} \\
\vspace*{0.3cm}
{\it $^3$ INFN, Sezione di Cagliari and Dipartimento di Fisica,  
Universit\`a di Cagliari,\\
C.P. 170, I-09042 Monserrato (CA), Italy} \\
\end{center}

\vspace{.1cm}

\begin{abstract}
A formalism proposed to study transverse $\Lambda$ polarization in unpolarized 
hadronic processes, based on a generalized pQCD factorization theorem, is 
extended to semi-inclusive DIS. Analytical expressions and examples of 
numerical estimates are given. 
\end{abstract}

\vspace{.7cm}

\renewcommand{\thefootnote}{\arabic{footnote}}
\setcounter{footnote}{0}

\nd 
{\bf 1. Introduction}
\nobreak
\vspace{6pt}
\nobreak

Transverse hyperon polarization in high energy, unpolarized
hadron-hadron collisions has formed a long-standing challenge for theoretical
models of hadronic reactions. The straightforward application of perturbative 
QCD and collinear factorization in the study of these observables is not 
successful, giving much too small values compared to the observed data, 
which may reach well above 20\%. 

Therefore, we have recently proposed a new approach \cite{abdm01} to this 
problem based on perturbative QCD and its factorization theorems, and which 
includes spin and intrinsic transverse momentum, $\bfk_\perp$, effects. 
It requires the introduction of a new type of leading twist fragmentation 
function (FF), which is polarization and $\bfk_\perp$-$\,$dependent, 
the so-called polarizing FF \cite{Mulders-Tangerman-96,abdm01}. 
Ideally, our approach could be tested by first extracting these new functions
in fitting some available experimental data, and then using the same functions
to give consistent predictions for other processes \cite{abdm02}. 
The problem with such a procedure is the actual availability of ``good''
experimental data, that is in kinematical regions appropriate to the 
application of our scheme.   

A similar approach, based on new polarization and $\bfk_\perp$-$\,$dependent
distribution and/or fragmentation functions, has already been applied to the 
study of transverse single spin asymmetries in inclusive pion production,
$\pup \, p \to \pi \, X$, at large $x_{_F}$ and medium to large 
$\bfp_{_T}$ \cite{abm95}.
 
An alternative, although somewhat related, approach, based on perturbative 
QCD and its factorization theorems with the inclusion of 
higher twist functions, has been investigated by Qiu and Sterman \cite{QS-91} 
and others.

The main idea behind the polarizing FF is that a transversely polarized
hyperon can result from the fragmentation of an {\em unpolarized\/} quark,
as long as the hyperon has a nonzero transverse momentum compared to the quark
direction (otherwise it would violate rotational invariance). This effect
need not be suppressed by inverse powers of a large energy scale, 
such as the center of mass energy $\sqrt{s}$ in $p\,$-$p$ scattering or 
the momentum transfer $Q$ in SIDIS. 

In our approach for the 
$p\,p \to \Lup \, X$ case, the transverse hyperon polarization in 
unpolarized hadronic reactions at large $p_{_T}$ can be 
written as follows \cite{abdm01}
\begin{eqnarray}
 &\,\,\,\,P_{_T}^{\,\Lambda}(x_{_F},p_{_T})&\, = \>\>
  \frac{d\sigma^{pp\to\,\Lambda^\uparrow\,X}-
   d\sigma^{pp\to\,\Lambda^\downarrow\,X}}
  {d\sigma^{pp\to\,\Lambda^\uparrow\,X}+
   d\sigma^{pp\to\,\Lambda^\downarrow\,X}} \label{ptlh}\\
 &\!\!\!\!\!\!\!\!\!\!\!\!\!\!\!\!\!\!\!\!\!\!\!\!=\!\!\!\!\!\!&
  \!\!\!\!\!\!\!\!\!\!\!\!\!\!\!\!\!\!\!\!\!
 \frac{\sum\,\int dx_a\,dx_b \int\! d^2\bfk_{\perp}\,
 f_{a/p}(x_a)\, f_{b/p}(x_b)\, d\hat\sigma(x_a,x_b;\bfk_{\perp})\,
 \Delta^{\!N}\!D_{\Lup\!/c}(z,\bfk_{\perp})}
 {\sum\,\int dx_a\,dx_b \int\! d^2\bfk_{\perp}\,
 f_{a/p}(x_a)\, f_{b/p}(x_b)\, d\hat\sigma(x_a,x_b;\bfk_{\perp})\,
 \hat D_{\Lambda/c}(z,\bfk_{\perp})}\,,\nonumber
\end{eqnarray}
where $d\sigma^{pp\to\Lambda\,X}$ stands for $E_{\Lambda}\,
d\sigma^{pp\to\Lambda\,X}/d^3\mbox{\boldmath $p$}_\Lambda$;
$f_{a/p}(x_a)$ and $f_{b/p}(x_b)$ are the usual unpolarized parton densities;
$d\hat\sigma(x_a,x_b;\bfk_{\perp})$ stands for 
$d\hat\sigma^{ab \to cd}/d\hat t$ and is the lowest order partonic 
cross-section with the inclusion of $\bfk_{\perp}$ effects; the $\sum$
takes into account all possible elementary interactions;    
$\hat D_{\Lambda/c}(z,\bfk_{\perp})$ and
$\Delta^{\!N}\!D_{\Lup\!/c}(z,\bfk_{\perp})$
are respectively the unpolarized and the polarizing 
FF \cite{abdm01,Mulders-Tangerman-96} for the process $c\to\Lambda+X$.
The polarizing FF is defined as: 
\bea 
\Delta^{\!N}\!D_{\hup\!/a}(z, \bfk_{\perp}) &\equiv& 
\hat D_{\hup\!/a}(z,\bfk_{\perp}) - \hat D_{\hdown\!/a}(z,\bfk_{\perp})  
\label{deld1}\nonumber \\ 
&=& \hat D_{\hup\!/a}(z,\bfk_{\perp})-\hat D_{\hup\!/a}(z,-\bfk_{\perp}) \>, 
\eea 
and denotes the difference between the density numbers  
$\hat D_{\hup\!/a}(z, \bfk_{\perp})$ and  
$\hat D_{\hdown\!/a}(z,$ $\bfk_{\perp})$ 
of spin 1/2 hadrons $h$, with longitudinal momentum fraction $z$, transverse  
momentum $\bfk_{\perp}$ and transverse polarization $\uparrow$ or  
$\downarrow$, inside a jet originated from the fragmentation of an  
unpolarized parton $a$. From the above definition it is clear that the 
$\bfk_{\perp}$ integral of the function vanishes and that, due to parity
invariance, the function itself vanishes in case the transverse momentum 
and transverse spin are parallel. Conversely for a hadron transversely
polarized along $\hat{\bfP}_h$, one can write
\be  
\hat D_{\hup\!/q}(z, \bfk_\perp) = \frac 12 \> \hat D_{h/q}(z, k_\perp) +  
\frac 12 \> \Delta^{\!N}\!D_{\hup\!/q}(z, k_\perp) \>  
\frac{\hat{\bfP}_h \cdot (\bfp_q \times \bfk_\perp)} 
{|\bfp_q \times \bfk_\perp|} \label{lamfn}
\ee
where $\hat D_{h/q}(z, k_\perp) = \hat D_{\hup\!/q}(z, \bfk_\perp) +   
\hat D_{\hdown\!/q}(z, \bfk_\perp)$ is the $k_\perp$-$\,$dependent 
unpolarized fragmentation function and  $k_\perp = |\bfk_\perp|$. 
We will adopt also the following notations:
\be
\Delta^{\!N}\!D_{\hup\!/q}(z, \bfk_\perp) \equiv 
\Delta^{\!N}\!D_{\hup\!/q}(z, k_\perp) \> 
\frac{\hat{\bfP}_h \cdot (\bfp_q \times \bfk_\perp)} 
{|\bfp_q \times \bfk_\perp|} = 
\Delta^{\!N}\!D_{\hup\!/q}(z, k_\perp) \> \sin\phi \>, 
\label{lamfn2}
\ee
where $\phi$ is the angle between $\bfk_\perp$ and $\bfP_h$, which,
in our configuration (as explained later), is the difference between
the azimuthal angles of $\bfP_h$ and $\bfk_\perp$, 
$\phi = \phi_{P_h} - \phi_{k_\perp}$.

Eq.~(\ref{ptlh}) is based on some simplifying assumptions (for a detailed
discussion we refer to \cite{abdm01}):
1) The $\Lambda$ polarization is assumed to be generated in the
fragmentation process; 2) We neglect the intrinsic 
$\mbox{\boldmath$k$}_\perp$ effects in the unpolarized initial nucleons; 
3)~The $\Lambda$ FF's also include $\Lambda$'s coming from decays of 
other hyperon resonances. Only  valence 
quarks in the polarized fragmentation process are considered.

Eq.~(\ref{ptlh}) was used in Ref.~\cite{abdm01}, together with a
very simple parameterization of the polarizing FF, to fit the available
existing data on $P^\Lambda_{_T}$. We consider now the same problem in SIDIS 
processes, $\ell \, p\to \ell' \, \Lup \, X$.
 
\vskip 18pt 
\nd 
{\bf 2. A Gaussian model for $\bfk_\perp$-$\,$dependent
fragmentation functions}
\label{gauss}
\vskip 6pt

We take into account only leading twist
and leading order contributions, looking at the process in the
virtual boson-target nucleon c.m. reference frame (VN frame).
Under these conditions the elementary virtual boson-quark scattering
is collinear (and along the direction of motion of the virtual boson)
and the transverse momentum of the final hadron with respect 
to the fragmenting quark, $\bfk_\perp$, coincides with the hadron
transverse momentum, $\bfp_{_T}$, as measured in the VN frame.

Due to the different kinematical scales involved in SIDIS \cite{abdm02}
we consider here a more realistic form of the polarizing FF than in 
Ref.~\cite{abdm01}, using for the unpolarized and polarizing FF 
the following general form:
\bea
 \hat{D}_{\Lambda/q}(z,\bfk_\perp) &=&
 \hat{D}_{\Lambda/q}(z,k_\perp) \> = \> 
 \frac{d(z)}{M^2}\,\exp\Bigl[\,-\frac{k_{\perp}^{\!\ 2}}
 {M^2 f(z)}\,\Bigr] \, ,
\label{defd} \\
 \Delta^{\!N}\!D_{\Lup\!/q}(z,k_\perp) &=&
 \frac{\delta(z)}{M^2}\,\frac{k_\perp}{M}\,
 \exp\Bigl[\,-\frac{k_{\perp}^{\!\ 2}}
 {M^2 \varphi(z)}\,\Bigr]\, ,
\label{defded}
\eea
where $M=1$ GeV/$c$ is a typical hadronic momentum scale and $f(z)$,
$\varphi(z)$ are generic functions of $z$, which we choose to be of
the form $Nz^a(1-z)^b$. 

Eqs. (\ref{defd}) and (\ref{defded}) must satisfy the positivity bound
\be
\frac{|\Delta^{\!N}\!D_{\Lup\!/q}(z,k_\perp)|}
{\hat{D}_{\Lambda/q}(z,k_\perp)} =
\frac{|\delta(z)|}{d(z)}\,\frac{k_\perp}{M}\,
\exp\Biggl[\,-\frac{k_\perp^{\!\ 2}}{M^2} \left( 
\frac{1}{\varphi}-\frac{1}{f} \right) \,\Biggr] \leq 1\>,
\label{posi}
\ee    
which, with  $\varphi(z) = r f(z)$, implies $r<1$ and 
\be
\label{pos2}
\frac{|\delta(z)|}{d(z)}  \leq   \left
[\frac{2\,e}{f(z)}\frac{1-r}{r}\right]^{1/2}
\>. 
\ee
The functions $d(z)$, $f(z)$ in Eq.~(\ref{defd}) are simply related to 
the usual, unpolarized and $\bfk_\perp$-$\,$integrated FF and to the hadron 
mean squared transverse momentum inside the observed fragmentation jet,
$\langle k_\perp^{\!\ 2}(z)\rangle$, giving for the unpolarized FF: 
\be
\label{dunp}
 \hat{D}_{\Lambda/q}(z,k_\perp) =
 \frac{D_{\Lambda/q}(z)}{\pi\,\langle k_\perp^{\!\ 2}(z)\rangle}\,
 \exp\Bigl[\,-\frac{k_\perp^{\!\ 2}}
 {\langle k_\perp^{\!\ 2}(z)\rangle}\,\Bigr]\, \cdot
\label{davek}
\ee
A comparison between Eqs. (\ref{davek}) and (\ref{defd}) yields the 
expressions of $d(z)$ and $f(z)$ in terms of $D(z)$ and 
$\langle k_\perp^{\!\ 2}(z)\rangle$.  
 
Some experimental information on $\langle k_\perp^{\!\ 2}(z)\rangle$ is 
available for pions but not yet for $\Lambda$ particles. 
To obey Eq.(\ref{pos2}) in a most natural and simple way we can write
\be
\frac{\delta(z)}{M^3} = \left[N_q \,\frac{z^\alpha (1-z)^\beta} 
{\alpha^\alpha \beta^\beta/(\alpha+\beta)^{(\alpha+\beta)}}\right]\,
\frac{D_{\Lambda/q}(z)}{\pi[\langle k_\perp^{\!\ 2}(z)\rangle]^{3/2}}\,
[2e(1-r)/r]^{1/2} \label{pardel} 
\ee
with $\alpha, \beta > 0$, $|N_q|\le 1$.

In this approach $\delta(z)$ is an unknown function depending on the  
parameters $\alpha$, $\beta$, $N_q$ and $r$, while $\varphi(z)$ is fixed by 
$r$, if we assume to know the functions $\langle k_\perp^{\!\ 2}(z)\rangle$
and $D_{\Lambda/q}$.

If one were to demand consistency with the results of
Ref. \cite{abdm01} $\delta(z)$ and $\varphi(z)$ could be fixed,
leading us to give predictions for $P^\Lambda_{_T}$ in SIDIS.
However, this would be an optimistic procedure, which could be
viewed only as a consistent example of using information
from one set of data, in order to give predictions for other
processes; it should be based on the assumption that all $p$-$p$
data originate from kinematical regions where pQCD and factorization
scheme hold, which is doubtful for the presently available
data.\footnote
{This procedure has been adopted in Ref.~\cite{abdm02}, which is
presently
under revision; we thank A.~Efremov, J.~Soffer and W.~Vogelsang for
emphasizing this point to us.}

In fact, recently it has become apparent that the formalism employed in
our $p$-$p$ paper \cite {abdm01} results in at most a few 
percents of the unpolarized cross-section, at least in the kinematical
regions where both polarization and cross-section data are available
(which is only a subset of the total region where data on
$P^\Lambda_{_T}$ have been published and used). This
casts doubts on the obtained polarizing FF and therefore these should not
be used to make predictions. A full assessment will be published
elsewhere.

We also notice that a similar situation holds for pion or
photon productions, in that the pQCD calculations, even in the central
rapidity region and at large $p_{_T}$ values, can be up to a factor 100 smaller
than data \cite{e706,ww,wang,zfpbl}; in those cases the discrepancy is
explained by the introduction of $\bfk_\perp$ effects in the
{\em distribution\/} functions: these give a large, spin independent,
enhancing factor, which brings the cross-sections in agreement with
data.
Such factors would not alter the calculation of the $\Lambda$
polarization in our approach, where spin effects are present
only in the fragmentation process, 
as they cancel in the ratio of cross-sections of Eq.
(\ref{ptlh}). However, it is too early to draw a definite conclusion,
and a more detailed study is in progress.

Here we only present the analytical formalism, which is meaningful and
valid in the appropriate regions, and show some numerical results
based on simple guesses for $N_q, \alpha, \beta, r$ on the basis of a
qualitative analysis of hadronic data and adopting the expression of
$\langle k_\perp^{\!\ 2}(z)\rangle$ valid for pions at LEP energies.

\vspace{16pt}
\goodbreak
\nd
{\bf 3. Analytical results}
\nobreak
\vspace{6pt}
\nobreak

We consider transverse $\Lambda$, $\bar\Lambda$ polarization in 
semi-inclusive DIS, both for neutral and charged current interactions 
(but neglecting electro-weak interference effects). We give the expression 
of the polarization as a function of the usual SIDIS variables, 
$x, y$, and $z_h$, in the current fragmentation region ($x_F>0$).

As already discussed, we present our results in the virtual 
boson-proton c.m. reference frame (VN frame). More specifically, we 
take the $\hat{z}$-axis of our frame along the direction of motion 
of the exchanged virtual boson; the $\hat{x}$-axis is chosen along the 
transverse momentum $\mbox{\boldmath$p$}_{_T}(=\bfk_\perp)$
of the observed hadron with respect to the $\hat{z}$-axis.
In this configuration, the transverse polarization is measured along 
the $+\hat{y}$-axis, and the angle $\phi$ is fixed at $\phi=\pi/2$. 

The transverse $\Lambda$ polarization is given by
\be
P_{_T}^{\,\Lambda}(x,y,z_h,p_{_T})=
\frac{d\sigma^{\Lambda^\uparrow}-d\sigma^{\Lambda^\downarrow}}
{d\sigma^{\Lambda^\uparrow}+d\sigma^{\Lambda^\downarrow}} \>,
\label{updow} 
\ee
with
\be
d\sigma^{\Lambda^{\uparrow(\downarrow)}}=
\frac{d\sigma^{\ell p\to\ell' \Lambda^{\uparrow(\downarrow)} X}}
{dx\,dy\,dz_h\,d^2\bfp_{_T}}=\sum_{i,j}\,f_{q_i/p}(x)\,
\frac{d\hat{\sigma}^{\ell q_i\to\ell' q_j}}{dy}
\hat{D}_{\Lambda^{\uparrow(\downarrow)}\!/q_j}(z_h,p_{_T}) \>,
\label{dgen}
\ee
where $i$,$j$ indicate different quark flavors and the sum includes
both quark and antiquark contributions. 

We consider separately different processes.

\goodbreak
\vskip6pt
\nd
{\bf a) $\ell^{\mp} \, p\to\ell^{\mp} \, \Lup \,X$}
\nobreak

This case is of interest for several experiments, {\it e.g.}
HERMES, H1 and ZEUS at DESY, COMPASS at CERN, E665 at SLAC.
One gets
\bea
P_{_T}^{\,\Lambda}(x,y,z_h,p_{_T}) &=&
\frac{\sum_q\,e_q^2\,f_{q/p}(x)\,\left[\,d\hat\sigma^{\ell q}/
dy\,\right]\,\Delta^{\!N}\!D_{\Lup\!/q}(z_h,p_{_T})}
{\sum_q\,e_q^2\,f_{q/p}(x)\,\left[\,d\hat\sigma^{\ell q}/
dy\,\right]\,\hat{D}_{\Lambda/q}(z_h,p_{_T})}\nonumber\\
&\simeq& \frac{(4u+d)\,\Delta^{\!N}\!D_{\Lup\!/u}
+ s\,\Delta^{\!N}\!D_{\Lup\!/s}}
{(4u+d)\,\hat{D}_{\Lambda/u}+s\,\hat{D}_{\Lambda/s}} \>,
\label{ncl}
\eea
where we have switched to the notation $f_{q/p}(x)\to q(x)$; the second
line is obtained by neglecting terms containing non-leading quark 
contributions both in the partonic distribution and fragmentation functions. 

A similar expression holds also for the 
$\ell^{\mp}\,p\to\ell^{\mp}\, \bar\Lambda^{\uparrow}\,X$ processes, with 
the exchange $q(x) \leftrightarrow \bar{q}(x)$.

\vskip6pt
\nd
{\bf b) $\nu\,p\to\nu\,\Lambda^\uparrow\,X$}

This process is of interest for the planned neutrino factories, and is 
currently under investigation by the NOMAD Collaboration at CERN. We get:
\be
P_{_T}^{\,\Lambda}(x,y,z_h,p_{_T}) \simeq
\frac{(Ku+d)\,\Delta^{\!N}\!D_{\Lup\!/u}+
s\,\Delta^{\!N}\!D_{\Lup\!/s}}
{(Ku+d)\,\hat{D}_{\Lambda/u}+s\,\hat{D}_{\Lambda/s}} \>,
\label{ncnu}
\ee
where $K=(1-8C)/(1-4C) \simeq 0.55$ (with $C=\sin^2\theta_{_W}/3 
\simeq 0.077$); terms quadratic in $C$ and non-leading quark contributions
have been neglected.

An analogous expression holds for $\bar\nu\,p\to\bar\nu\,\Lup\,X$ 
processes, with a factor $K$ ranging now between 0.78 and 4 for $y=0$ and 
$y=1$ respectively.

\vskip6pt
\nd
{\bf c) $\nu\,p\to\ell^-\,\Lambda^\uparrow\,X$}

This case is again of interest for neutrino factories and for the NOMAD 
experiment, which recently have published results for
$\Lambda$ and $\bar{\Lambda}$ polarization \cite{nomad00,nomad01}.  
One~finds
\be
P_{_T}^{\,\Lambda}(x,y,z_h,p_{_T})=\frac{(d+R\,s)\,
\Delta^{\!N}\!D_{\Lup\!/u}+\bar{u}\,
(\Delta^{\!N}\!D_{\Lup\!/\bar{d}}+
R\,\Delta^{\!N}\!D_{\Lup\!/\bar{s}})\,
(1-y)^2}{(d+R\,s)\,\hat{D}_{\Lambda/u}+\bar{u}\,
(\hat{D}_{\Lambda/\bar{d}}+R\,\hat{D}_{\Lambda/\bar{s}})\,
(1-y)^2} \>,
\label{ccnu}
\ee
where $R=\tan^2\theta_C\simeq0.056$. Neglecting sea-quark 
contributions leads to 
\be
P_{_T}^{\,\Lambda}(x,y,z_h,p_{_T})\simeq
\frac{\Delta^{\!N}\!D_{\Lup\!/u}}
{\hat{D}_{\Lambda/u}} \>\cdot
\label{ccnusim}
\ee
The same expression is true for the case $\ell^+\,p\to\bar\nu\,\Lup\,X$.

\vskip6pt
\nd
{\bf d) $\bar{\nu}\,p\to\ell^+\,\Lambda^\uparrow\,X$}

This case is very close to the previous one, with obvious
modifications:
\be
P_{_T}^{\,\Lambda}(x,y,z_h,p_{_T})=
 \frac{(1-y)^2\,u\,(\Delta^{\!N}\!D_{\Lup\!/d}+
 R\,\Delta^{\!N}\!D_{\Lup\!/s})+
 (\bar{d}+R\,\bar{s})\,\Delta^{\!N}\!D_{\Lup\!/\bar{u}}}
 {(1-y)^2\,u\,(\hat{D}_{\Lambda/d}+R\,\hat{D}_{\Lambda/s})+
 (\bar{d}+R\,\bar{s})\,\hat{D}_{\Lambda/\bar{u}}} \>\cdot
\label{ccnub}
\ee
Again, when sea quark contributions are neglected and isospin
symmetry is invoked, we find the very simple expression
\be
P_{_T}^{\,\Lambda}(x,y,z_h,p_{_T})\simeq
\frac{\Delta^{\!N}\!D_{\Lup\!/u}+
R\,\Delta^{\!N}\!D_{\Lup\!/s}}
{\hat{D}_{\Lambda/u}+R\,\hat{D}_{\Lambda/s}} \>\cdot
\label{ccnubsim}
\ee
The same expression holds for the $\ell^-\,p\to\nu\,\Lup\,X$ case.

The above results, Eqs. (\ref{ncl})-(\ref{ccnubsim}), relate measurable
polarizations to different combinations of (known) distribution functions,
(less known) unpolarized and (unknown) polarizing FF; the different terms 
have relative coefficients which depend on the dynamics of the elementary 
partonic process and/or on the relevance of $s$ quark contributions
in the partonic distribution functions. 

This large variety of possibilities gives a good opportunity to investigate 
and test the relevant properties of the unpolarized and polarizing $\Lambda$
FF, by measuring the hyperon transverse polarization. In some special 
cases, Eq. (\ref{ccnusim}), experiment offers direct information on these 
new functions. 

Analogous results hold for the production of $\bar\Lambda$ \cite{abdm02}.

\vspace{16pt}
\goodbreak
\nd
{\bf 4. Examples of numerical estimates}
\nobreak
\vspace{6pt}
\nobreak

Let us consider the parameters $\alpha, \beta, N_q, r$ appearing in Eqs.
(\ref{defded}) and (\ref{pardel}), and their possible values.
Looking at the data on transverse $\Lambda$ polarization in hadronic
reactions we expect: negative contributions for up and down quarks 
($N_{u,d} < 0$)\footnote{Isospin symmetry is assumed to hold, that is we
take $(\Delta^{\!N})D_{\Lambda/d} = (\Delta^{\!N})D_{\Lambda/u}$.}
and positive for strange quarks ($N_s > 0$), in order to have 
$P^\Lambda <0$ and $P^{\bar\Lambda}\simeq 0$; a polarizing FF peaked at 
large $z$ to explain the increasing in magnitude of the polarization 
with $x_F$ at
fixed $p_T$ (thus implying large values of $\alpha$, while $\beta \simeq 
O(1)$); a Gaussian shape similar for unpolarized and polarizing FF to 
explain the large values of the polarization (which means $r\simeq O(1)$). 

We then fix $\beta = 1$, $r = 0.7$; 
$\sqrt{\langle k_\perp^{\!\ 2}(z)\rangle}=0.61\, z^{0.27}(1-z)^{0.2}$
GeV/$c$ (as used in Ref.~\cite{abm95}).
We adopt for the unpolarized, $\bfk_\perp$-$\,$integrated $\Lambda$ FF, 
the $SU(3)$ symmetric parameterizations of Ref.~\cite{boros2}, and 
consider for the polarizing FF:
$1)$ a scenario with almost the same weight for up and strange quarks, 
with $N_u=-0.8$ and $N_s=1$ ($\alpha \simeq 6$)
and $2)$ a scenario similar to the model of Burkardt and Jaffe \cite{buja93}  
for the longitudinally polarized FF $\Delta D_{\Lambda/q}$, with
$N_u=-0.3$ and $N_s=1$ ($\alpha \simeq 4$). 

We consider kinematical configurations typical of running experiments
(HERMES at DESY, NOMAD at CERN, E655 at SLAC). 

Since the $Q^2$ evolution of the polarizing FF is not under control
at present, and the HERMES and NOMAD experiments involve a
relatively limited range of $Q^2$ values, in our numerical
calculations we have chosen a fixed scale, $Q^2=2$ (GeV/$c)^2$. 
For the unpolarized distribution functions
we have adopted the MRST99 set \cite{mrst99}. 

\begin{figure}[htbp] 
\begin{center}
\hspace*{-0.2cm}
\mbox{~\epsfig{file=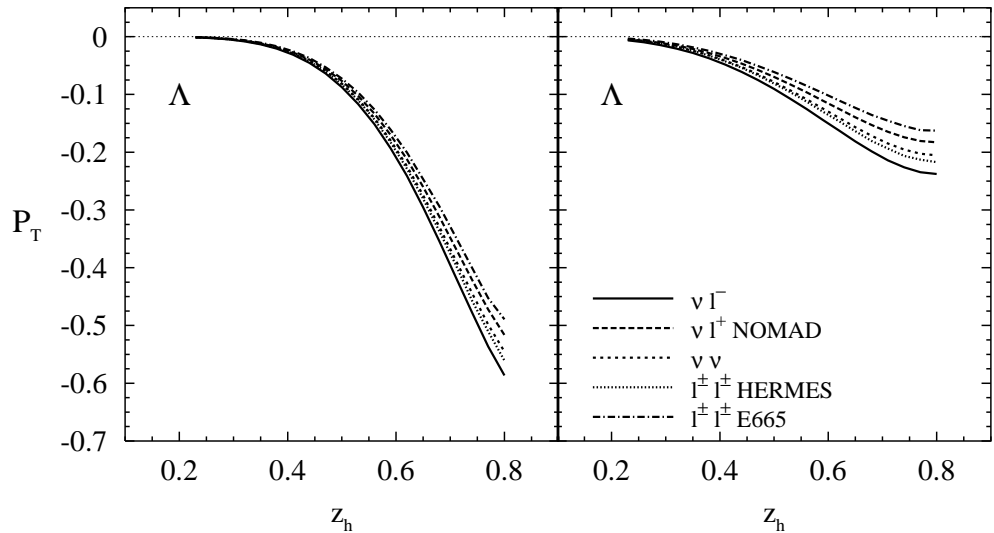,angle=-90,width=12.8cm}}
\end{center}
\caption{\label{pla}
{\footnotesize Transverse $\Lambda$ polarization, 
$P_{_T}^{\Lambda}$, {\it vs.} $z_h$ and
averaged over $p_{_T}$, for several SIDIS production processes, 
with scenario 1 (on the left) and scenario 2 (on the right) 
for the polarizing FF.}}
\end{figure}

\begin{figure}[hbtp] 
\begin{center}
\hspace*{-0.2cm}
\mbox{~\epsfig{file=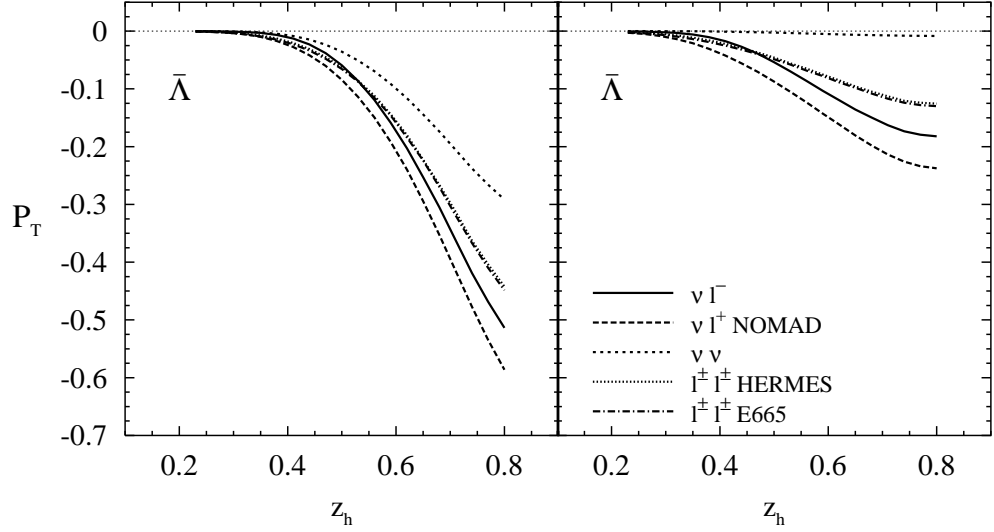,angle=-90,width=12.8cm}}
\end{center}
\caption{\label{plb}
{\footnotesize  Transverse $\bar\Lambda$ polarization, 
$P_{_T}^{\bar\Lambda}$, {\it vs.} $z_h$ and
averaged over $p_{_T}$, for several SIDIS production processes, 
with scenario 1 (on the left) and scenario 2 (on the right) 
for the polarizing FF.}}
\end{figure}

Our results are shown in Figs.~\ref{pla}-\ref{plb}. Fig.~\ref{pla} shows 
$P_{_T}^{\Lambda}$ as a function of $z_h$, after $p_{_T}$ average,
for all the cases {\bf a)-d)} considered above and for kinematical 
configurations typical of the corresponding relevant experiments (as 
indicated in the legends); the two plots (left and right) differ by the 
scenario of the polarizing FF.

The polarization is in general large in magnitude and negative:
contributions from $\Delta^{\!N}\!D_{\Lup\!/s}$ 
are always suppressed, either by the $s$ quark 
distribution [via factors like $s/(Ku + d)$] or by the Standard Model 
factor $R$, see Eqs. (\ref{ncl})-(\ref{ccnubsim}). Thus, the strange quark 
contribution is suppressed, unless one uses a $SU(3)$ asymmetric FF set
for which $|\Delta^{\!N}\!D_{\Lup\!/s}| \gg |\Delta^{\!N}\!D_{\Lup\!/u,d}|$. 
This is reflected by the fact that in Fig. \ref{pla} 
all processes have similar polarizations, approximately given by 
the $p_{_T}$-averaged ratio $\Delta^{\!N}\!D_{\Lup\!/u}/\hat{D}_{\Lambda/u}$. 
The polarization is smaller in the right plot (scenario 2) simply because
$|\Delta^{\!N}\!D_{\Lup\!/u}|$ is smaller.

Fig.~\ref{plb} shows the corresponding results for the case of 
$\bar\Lambda$ SIDIS production; here the effect of
cancellations between up (down) and strange contributions 
is more significant: notice, for instance, how 
$P_{_T}(\nu\,p\to\nu\,\bar\Lambda^{\uparrow}\,X)$ is suppressed 
compared to the analogous process for $\Lambda$. 

\vspace*{-.3cm}


In conclusion, single spin effects, 
suppressed in leading twist collinear applications 
of pQCD factorization theorems, may instead reveal new interesting aspects 
of non-perturbative QCD. Here we have considered the long-standing problem 
of the transverse polarization of hyperons produced from {\it unpolarized}
initial nucleons. Our project is based on the use of a QCD factorization 
scheme, generalized to include intrinsic $\bfk_\perp$ in the
fragmentation process: this allows to introduce new spin dependences
in the fragmentation functions of unpolarized quarks.

These new functions, the polarizing FF, are supposed to describe universal 
features of the hadronization process, which is factorized in a similar
way as the usual $\bfk_\perp$-$\,$integrated FF.
If correct, this idea should allow a consistent phenomenological description 
of hyperon polarization in different processes, provided data are in 
kinematical regions where a hard scattering approach can be relevant.

\vspace{16pt}
\goodbreak
\nd
{\bf Acknowledgements}
\nobreak
\vspace{6pt}
\nobreak

M.A. and U.D. would like to thank the organizers  
for their kind invitation to such an extremely interesting Conference. 
U.D. and F.M. are grateful to CO\-FI\-NAN\-ZIAMENTO MURST-PRIN 
for partial support.

\end{document}